# Beam Instrumentation

*Raymond Veness*
CERN, Geneva, Switzerland

**Abstract**
Beam instrumentation is known as the 'eyes and ears of a particle accelerator'. It provides the data required for the most basic operating functions such as beam steering and intensity measurement, but also for diagnosing problems and optimising performance. This paper describes some of the many techniques used in the field, with an emphasis on mechanical designs, issues and current research trends.

## 1 Introduction

The term 'beam instrumentation' (or also beam diagnostics) covers the systems used to provide data derived from the particle beam used for the operation of an accelerator. This scope therefore excludes the particles detectors used for experimental physics as well as the multitude of instruments used for monitoring the operation of other accelerator components (eg, flow meters for cooling systems).

The needs for instrumentation in an accelerator are varied, but can be grouped into three categories:

- keeping beams on-track, either through measuring the beam position as it orbits an accelerator ring, or for steering the beams through a transfer line or LINAC,
- ensuring safe operations, observing beam losses (often detecting secondary particles created when part of the accelerator beam is lost) which can cause damage or irradiation of equipment,
- optimising performance and diagnosing performance issues.

The types of instruments, from the physical principles used for measurement, through the engineering design and acquisition techniques also vary widely. Different particle types, with different electric charges, mass, intensities and energies typically require different types of instruments. In addition, the different measurement needs result in very different engineering designs – for example, setting-up the H- beam in the CERN LINAC4 requires a completely different approach to monitoring the beam size in the high energy LHC ring.

Physical principles for beam instrumentation, performance as well as the physics of the beam-matter interactions relating to instrument performance are well covered in other CAS schools [1-3]. This paper will focus on the mechanical engineering design principles, materials, components and challenges involved in fulfilling some of these requirements. In addition, beam instruments are complex engineering objects, requiring the skills of most of the lectures presented in this school. This paper will therefore also avoid reiterating the many key accelerator engineering fundamentals presented in the earlier lectures.

## 2 Movement

The majority of instruments used in an accelerator are ‘static’. They use principles such as electro-magnetism or ionisation of residual gas to interact with the beam and then passively detect a signal. However, a significant fraction of instruments needs to move within the vacuum envelope of the accelerator. There are typically three reasons for this:

- the instrument has enough material to be ‘destructive’ to the beam and is used for setting-up beamlines then removed for operations,
- the instrument needs to approach the beam to improve performance, for example when the beam size decreases during the acceleration cycle,
- the instrument needs to ‘scan’ the beam by moving some material (or even a laser beam) across the beam to cause some deliberate interaction or loss which can be measured as a function of the material position.

Set-up instruments typically use an ‘in-out’ type movement with two fixed positions. This movement can be made by simple pneumatic cylinders. The positions are normally ‘interlocked’ with end switches to avoid damage if the instruments are not removed for operations. For more critical locations, the movement is also fitted with a spring to ensure that the instrument is retracted in case of power loss.

For instruments in the other two categories where measurement of the intermediate position is important then the movement is normally made with a ball screw drive and the position if determined either with a system to count the turns of the screw such as a stepper motor or a magnet resolver. Alternatively, an independent linear measure such as a LVDT or optical ruler can be used.

## 3 Transmission

### 3.1 Transmission of movement

Existing instrumentation in use requires the movement system to be outside of the vacuum envelope and so require a transmission of movement from air to vacuum. Development is in progress for future systems to use fully UHV compatible linear drives such that only electrical power and position signals need to be transmitted.

The current state-of-the-art for movement transmission is by using magnetic coupling. This can be directly using rotary movement, such as in the LIU wire scanner where a permanent magnet synchronous motor (PMSM) is separated into an in-vacuum rotor and an ex-vacuum stator by a thin-walled stainless steel sleeve (see Fig. 1). Alternatively, magnetically coupled linear drives have been developed and commercially available. These are successfully used for in-out movement drives such as the CERN injector BTV (beam “television” observation screen) instrument (see Fig. 2).

For the remaining instruments, either in more classical systems or where a very direct control through vacuum is required, edge-welded bellows are widely used. These have the advantage of a long ratio of travel to free length and a low stiffness. There are a number of commercial suppliers which produce designs for ‘standard’ geometries which are well established and tested. In recent years reliability issues have been seen in accelerator devices with edge-welded bellows, in particular for unconventional operating conditions such as fast movements or transverse (shear) movements. Recent studies on these failures have led to an understanding of the inherent complexity of an analytical design of such complex objects. A complete simulation needs to take into account elastic-plastic, work-hardening material in a heat-affected zone with both low and high-cycle fatigue.

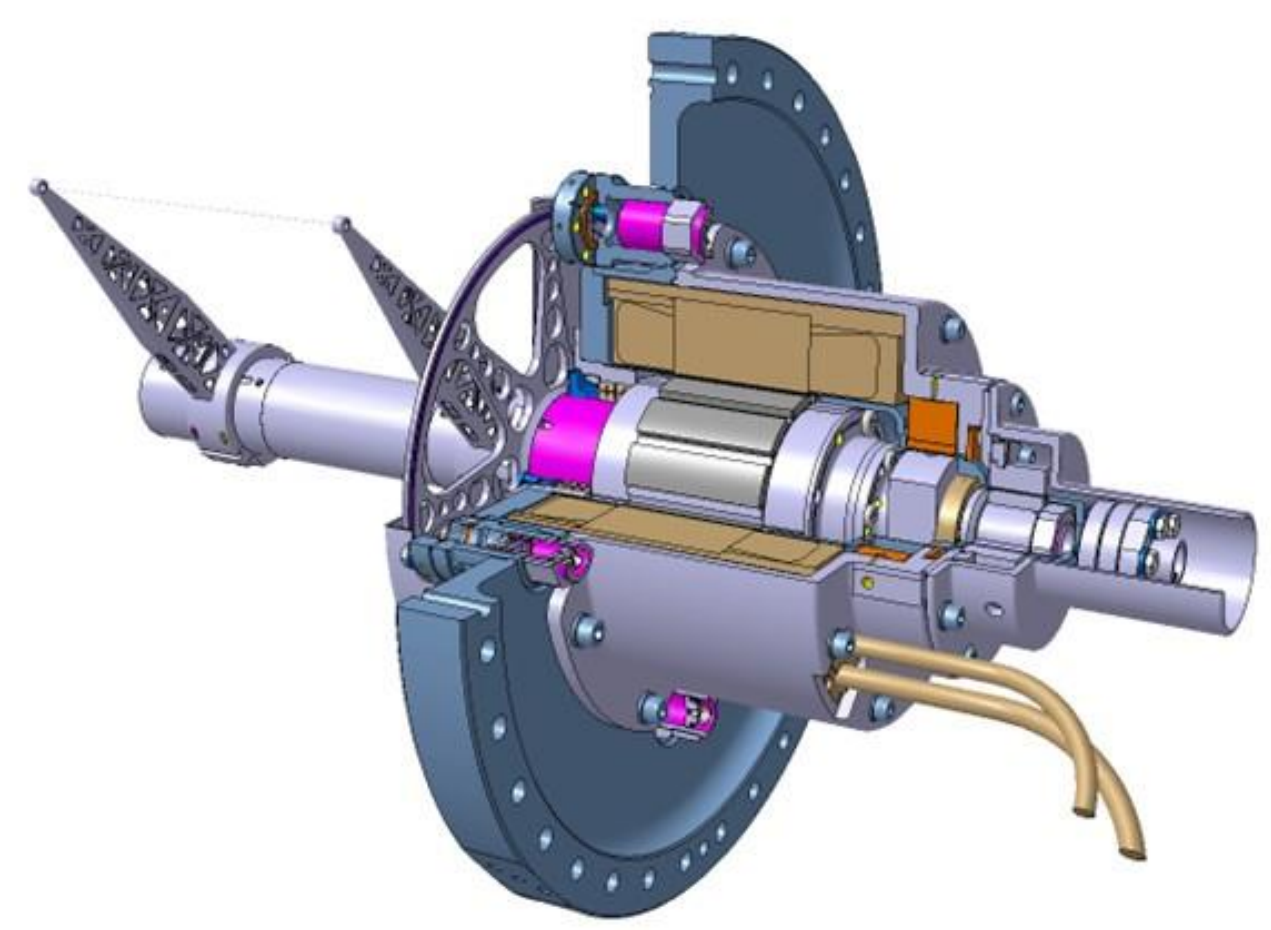

**Fig. 1:** LIU wire scanner showing magnetically coupled motor drive.

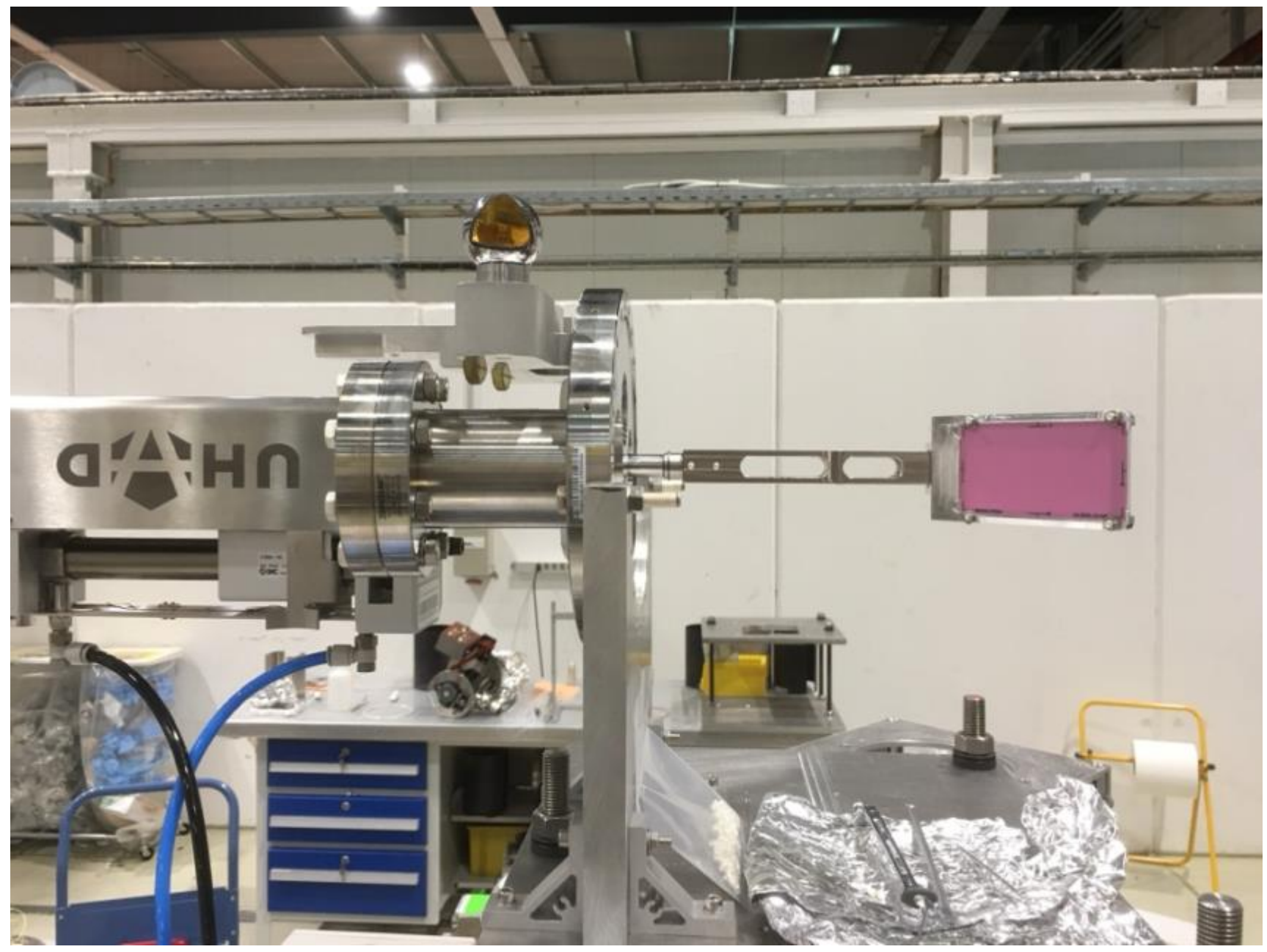

**Fig. 2:** BTV instrument showing magnetically coupled push-pull transmission.

### 3.2 Electrical feedthroughs

Beam instrumentation typically only require low voltage signal transmission into vacuum. Commercial multi-pin connectors can be used. In instruments such as SEM grids, the number of channels can be limited by available space for multi-channel feedthroughs. Typical reliability issues relate to poor material quality from suppliers and vacuum leaks after brazing or welding of ceramics into flanges.

A notable challenge for instrument feedthroughs is for position pickups, particularly in cryogenic applications. Measurements require impedance-matched coaxial RF cables and feedthroughs. These must also be vacuum tight to avoid virtual leaks in insulation vacuum [4]. Manufacturers are few and designs are specific to the pickup design.

### 3.3 Optical viewports

Optical viewports are generally avoided in accelerator design due to their relative fragility and complexity. However, they are often necessary in beam instrumentation. Applications include: direct observation of scintillating or other light-producing screens (called BTVs or beam televisions at CERN); transmission of synchrotron radiation (SR) into dedicated SR diagnostics; transmission of laser light, used for non-contact position measurement (interferometry or optical rulers). Viewports are also used to verify the integrity of delicate instruments such as wire scanners.

The application will determine both the viewport aperture and its material. Fused silica, borosilicate glass, sapphire may be specified depending on optical requirements as well as coatings for transmission and reflectivity. Viewports are normally procured or assembled ex-situ onto vacuum flanges of standard sizes. Design and manufacture are specialized skills with a number of challenges. Making a leak-tight joint with a glass normally requires brazing for UHV tightness. Coefficient of thermal expansion (CTE) of window materials are typically small and not matched to most metals. A requirement for bakeability makes designs more challenging. Designs (see Fig. 3) therefore normally consist of a transition piece in a metal with an intermediate CTE (eg, Kovar) which is brazed to the window material. This transition should be as flexible as possible to compensate for differential expansion. This sub-assembly is then welded into the vacuum flange.

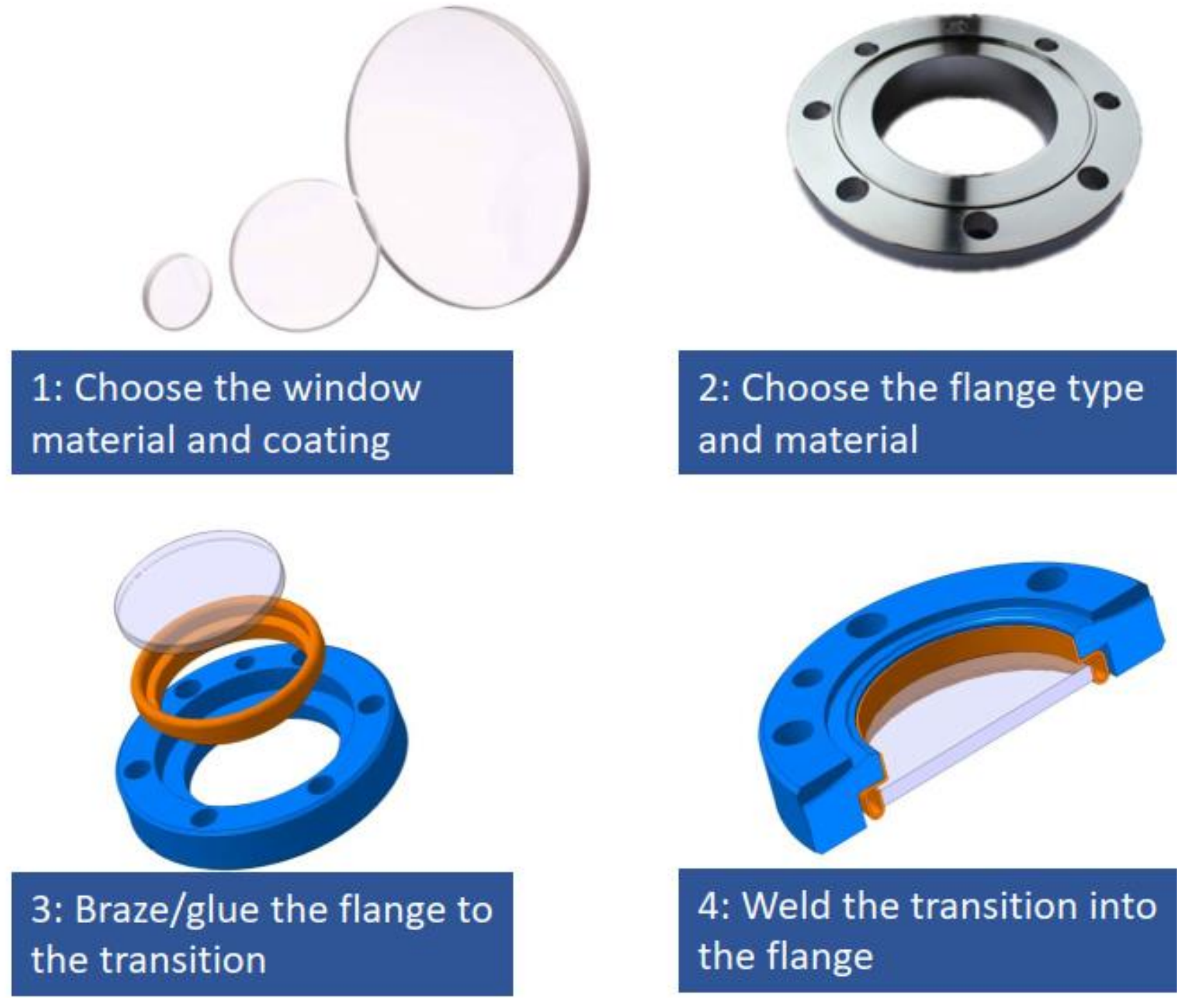

**Fig. 3:** Components of a UHV viewport.

Viewports are fragile and typically represent a single point of failure for the accelerator. Most, if not all accelerator complexes have experienced accidental venting due to viewport failure. Design and quality assurance are critical. However, there are a number of manufacturers in the world each with proprietary designs and materials as well as many stockists, sometimes selling different diameters from different manufacturers. Failures commonly occur during assembly, due to flange deformation or during bakeout. However, viewport materials are generally brittle and prone to cyclic crack growth with fast fracture failures occurring after a number of cycles of venting or bakeout. Regular inspection is therefore important.

# 4 Instrument-beam interface

## 4.1 Beam intercepting materials

There are several applications for beam intercepting materials in accelerators, in particular targets and beam dumps which are covered in other parts of this school. There are also widely used engineering tools for simulating beam-material interactions such as Fluka and GEANT as well as Finite Element (FE) tools for thermo-mechanical simulations. We will focus here on applications for instrumentation.

There are two broad categories of instruments that use a physical interaction between solid materials and beams:

### *4.1.1 Observation screens*

These intercept the beam with a thin sheet of material which produces light by, for example, scintillation. Functional screen materials can be ceramics or metals which, from an engineering perspective must be vacuum compatible and sufficiently resistant to temperature and thermal shock. Use tends to be limited to lower energies and intensities due to the larger material interactions but research for new materials is in progress.

### *4.1.2 Wires and foils*

Usually metal, intercept the beam and use either detect a flow of electrical current in the material caused by the emission of secondary electrons (called SEM wires/grids/foils) or by locally scattering some beam particles which are then detected downstream by dedicated beam-loss instruments. A number of materials are used, for example, carbon, tungsten, beryllium, boron, driven primarily by beam energy and intensity to give the right signal, but then by thermal resistance, mechanical properties and availability in the right dimensions.

These are also limited by material damage, but use of low atomic number (Z) materials, small diameters/thicknesses (down to the 10's of um) and rapid movement of scanning instruments makes them still partly useable for the latest high energy machines (see Section 6.1.2). Another limitation for superconducting machines can be the energy deposited by interactive beam loss in downstream cryo-magnets. Again, this is an active subject for research in instrumentation mechanics.

## 4.2 Beam-coupled impedance heating

Many instruments require either a change in the beam pipe aperture or have some structure inside the beam vacuum. Such changes can cause resonances in the electro-magnetic field which accompanies the circulating beam similar to a radio antenna. These can cause local heat deposition and heating in the instrument as well as potentially beam instabilities.

Simulating these resonances and their impact on the instrument is a specialist skill which has seen much progress in the last years. Figure 4 shows a simulation of a beam spectrum in the SPS at CERN (injection and extraction energies in colours) overlaid with the simulated resonances of an instrument (LIU wire scanner). It can be seen that the resonances can be very 'narrow band' with orders of magnitude of power dissipation between on and off resonance so there is significant potential for heating damage in high intensity circulating machines. As a general design rule, avoid rapid geometry changes wherever possible. It is important to involve simulation experts early in an instrument design phase. However, designing for the worst-case resonances can be expensive or severely limit the instrument performance.

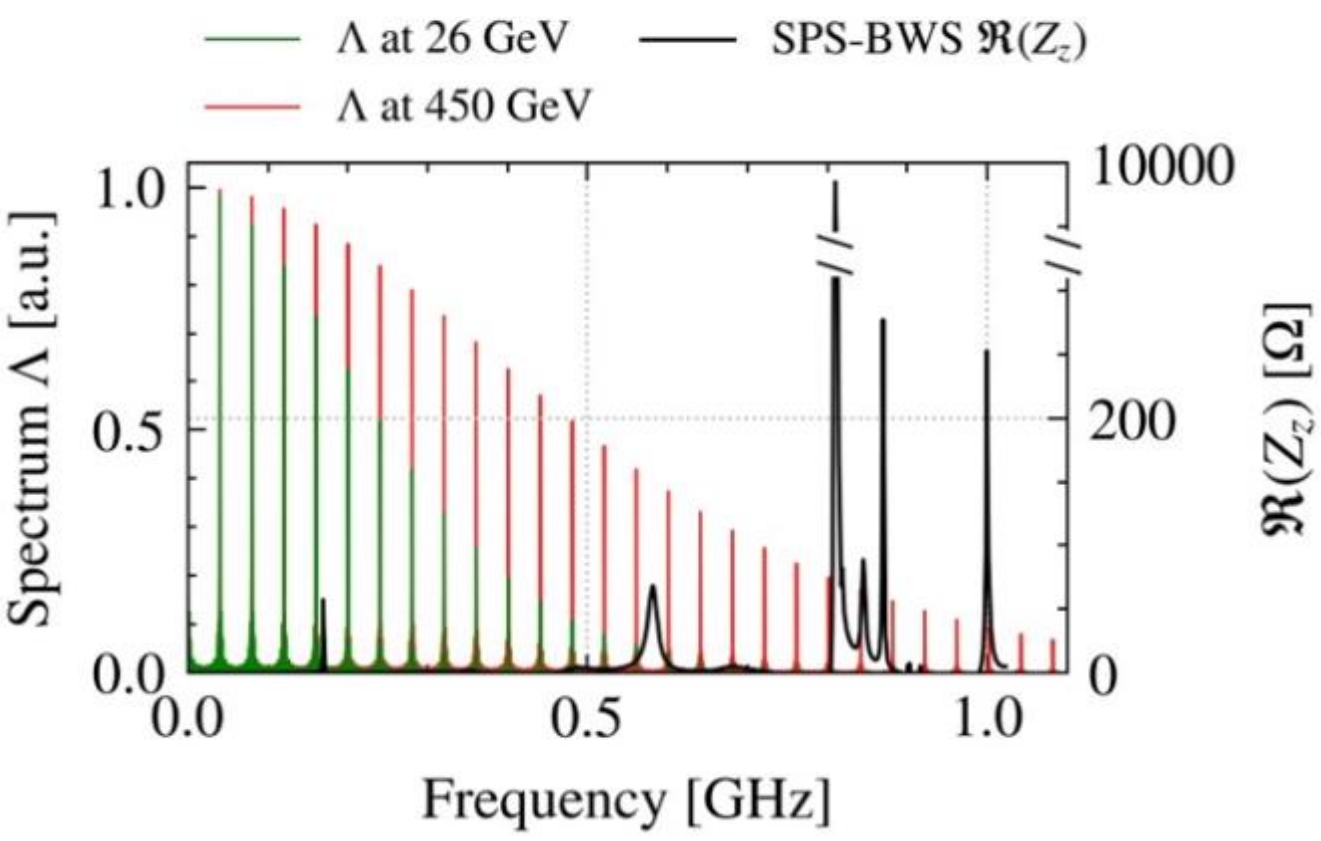

**Fig. 4:** Beam spectrum in the SPS (green) overlaid with wire scanner resonant modes (black), from Ref. [5].

### 4.3 Heat transfer under vacuum

As seen in Section 4.2, it can be difficult to accurately predict the heat deposited on in-vacuum components close to the beam so effective heat removal can be an important design factor. However, heat transfer under vacuum is non-trivial. There is no convection under vacuum. Heat conduction between bolted surfaces is much lower (sometimes orders of magnitude) than in air as conduction between non-continuous surfaces in air is assisted by micro-convection. Relying on conduction under vacuum requires careful design, simulation and perhaps test.

This leaves heat transfer by thermal radiation, which can be reliably simulated. However, from the Boltzmann equation, (Eq. (1)), it can be seen that the power transferred is proportional to the 4$^{th}$ power of the absolute temperature differences between the surfaces, so only starts to become effective at large temperature differences. For example, a 5x5 cm surface radiating to a room temperature vacuum chamber will emit just 0.5 W at 50 °C but 16 W at 300 °C. So, in summary, even relatively small heat loads from, for example, beam-coupled impedance, can cause steady-state temperatures in the 100s of °C and consequent damage to equipment. An example is shown in Fig. 5 where a carbon scanner wire failed by sublimation (so 1000+ °C) out-of-beam in the SPS.

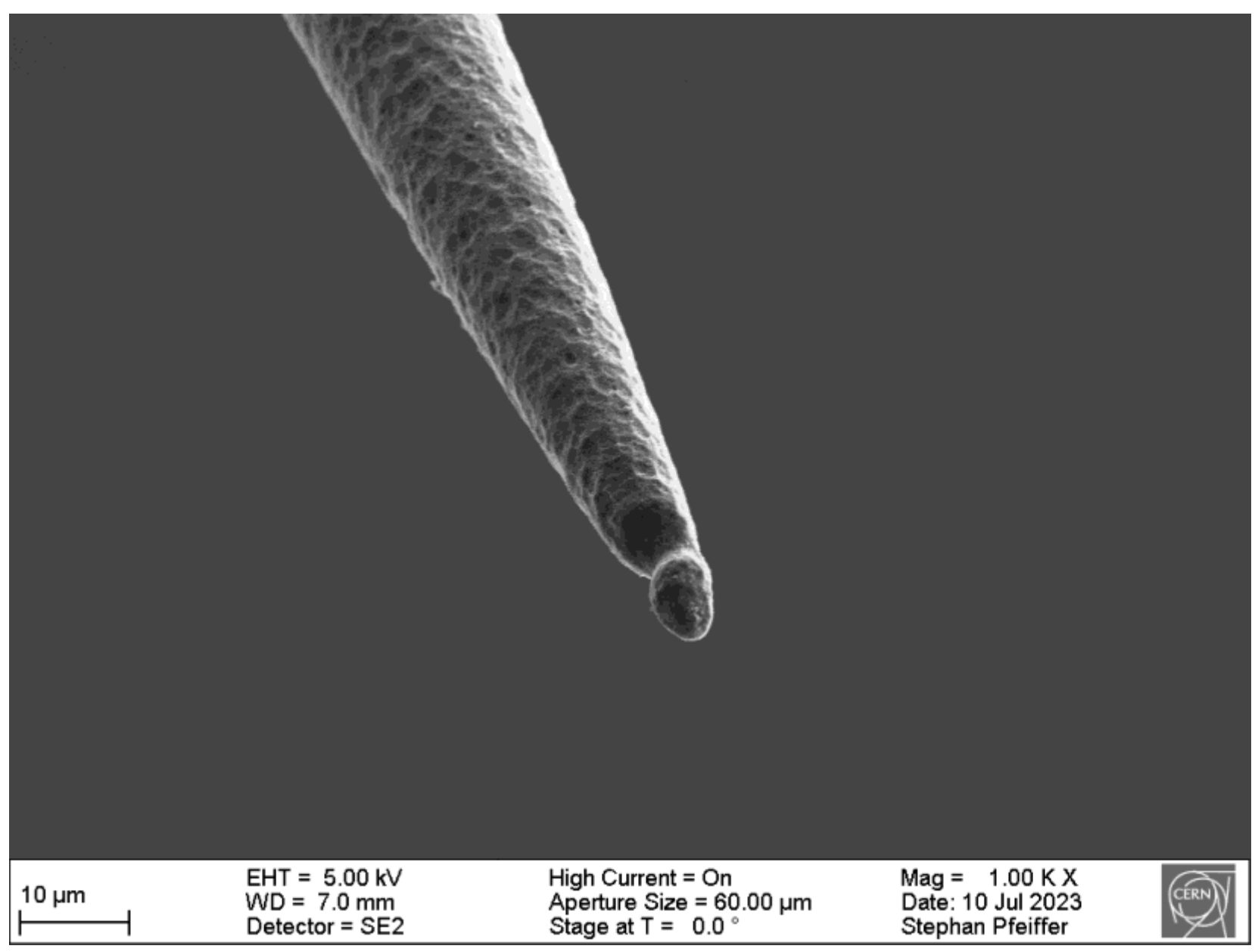

**Fig. 5:** Micrograph of sublimated scanner wire (courtesy CERN EN/MME).

There are a number of solutions to these problems:

- avoid by design, for example making instruments retractable and/or hidden behind RF screens (see Fig. 6), or using materials that can resist very high temperatures,
- add RF-absorbing materials such as ferrites, or introduce couplers to extract power and dampen specific resonant frequencies,
- include cooling, either active (eg, water) or carefully designed conduction paths,
- include temperature monitors where possible.

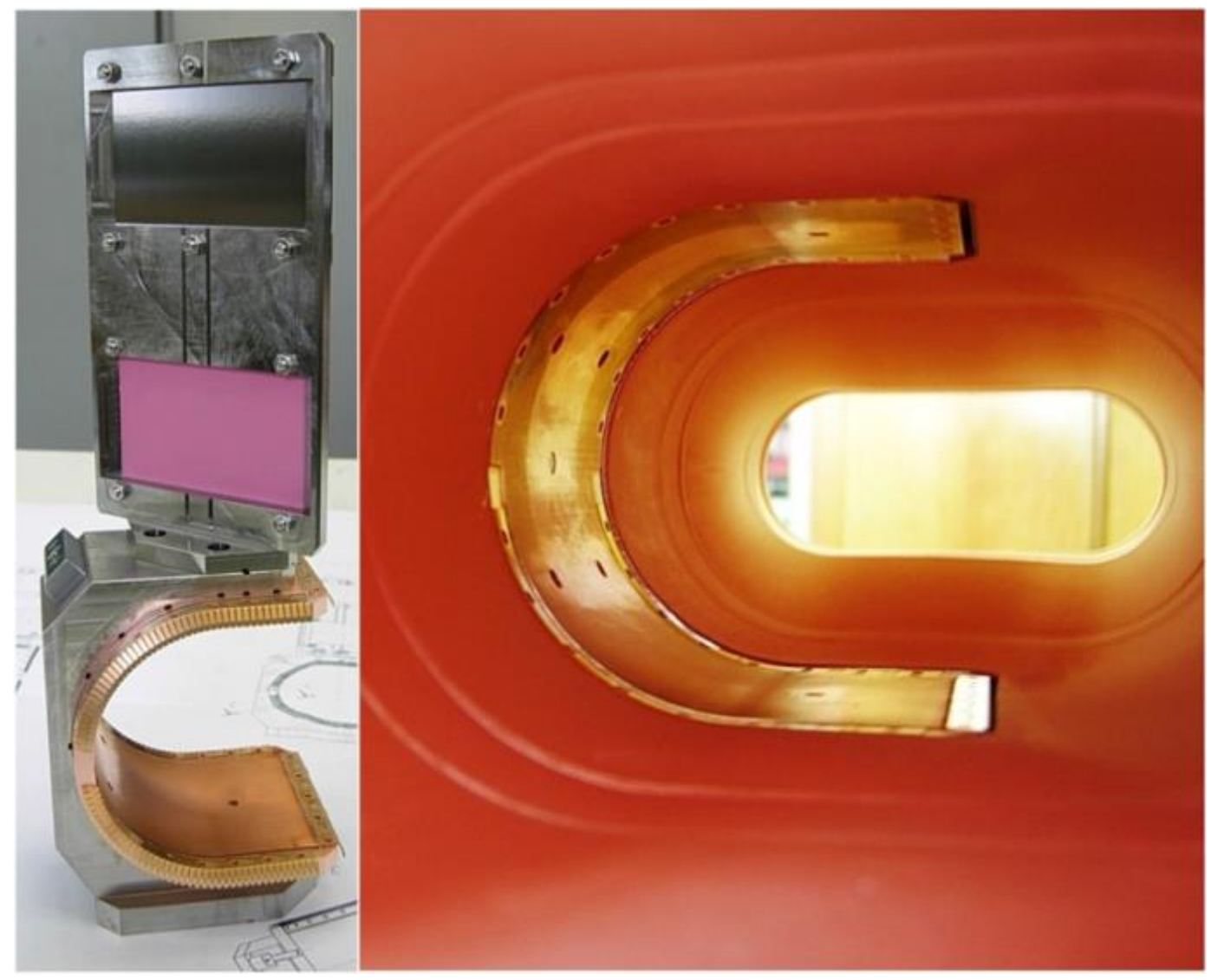

**Fig. 6:** LHC injection BTV showing impedance minimisation design.

## 5 Advanced materials and production processes

### 5.1 Low-density materials

As seen in Section 4, interceptive instruments are limited by damage due to beam-induced heating. New developments into nano-structured materials such as graphene and Carbon Nano-Tubes (CNTs) are opening new avenues to extend the reach of these instruments to higher energy and intensity machines.

Taking the example of wires for grids and scanners, key physical properties are:

- low density/atomic number (intercepting less beam),
- high mechanical strength (requiring less material),
- high melting/sublimation temperature (tolerating higher deposited energies),
- high heat capacity (giving lower temperatures for the same deposited energy).

Classical materials selection techniques such as the ‘Ashby plot’ allow potential materials to be categorised according to these criteria. Figure 7 [6] shows such a plot for these criteria. It is clear that low-density carbon-based materials have a high potential for extending the reach of both 2D (wire) and 3D (screen) interceptive detectors.

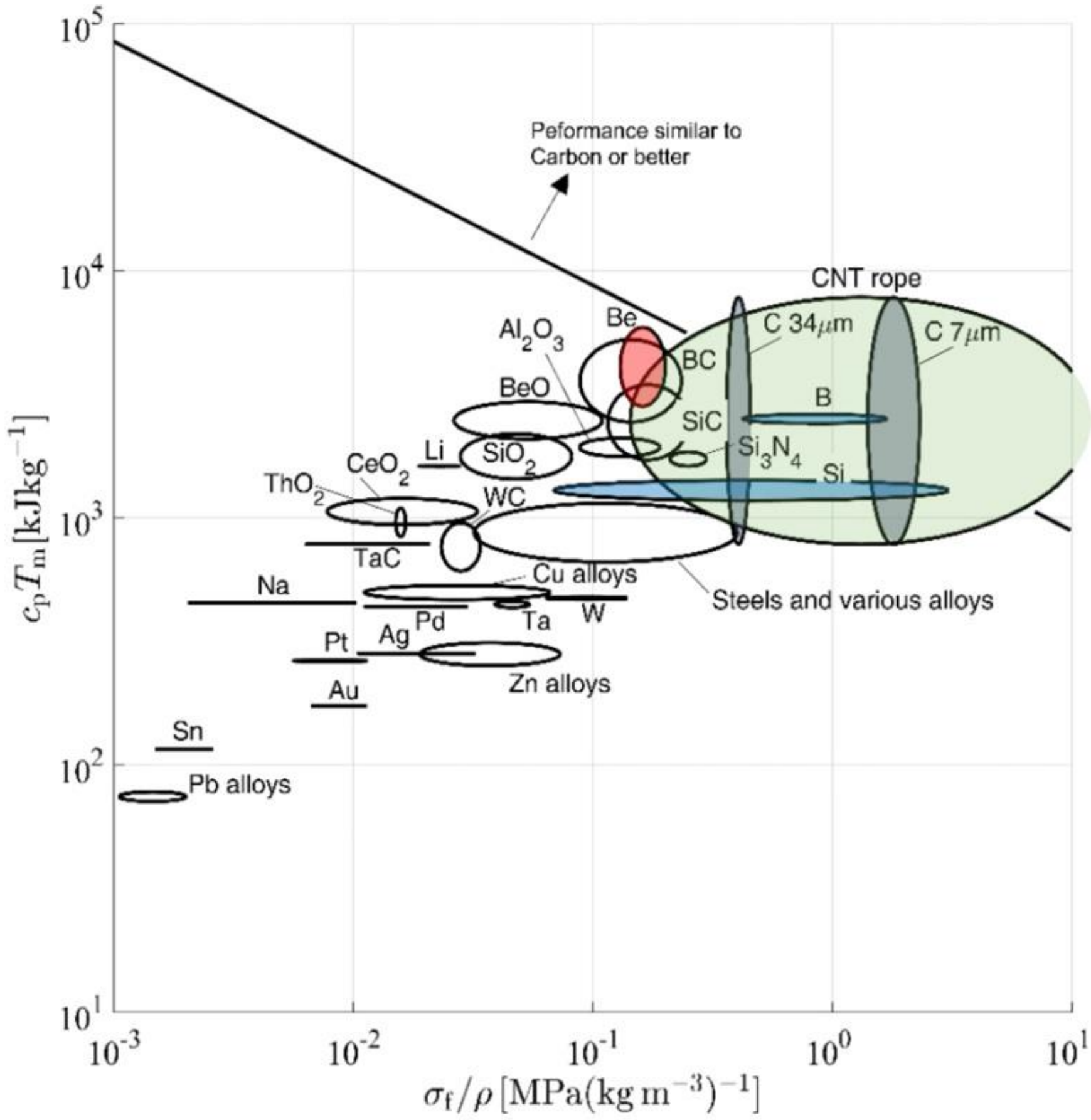

**Fig. 7:** Ashby plot classifying potential materials for a beam intercepting instrument.

Considering 2D wires, individual CNTs are only commercially available in sub-mm lengths. Longer structures called wires or ropes are made by twisting many shorter CNTs. These ropes already have physical properties more interesting than carbon fibres, but are orders of magnitude below single fibre limits.

Active research is ongoing with potential for significant increase in performance for these intercepting detectors [7], with beam tests both in facilities such as HiRadMat at CERN and in intercepting instruments at PSI and CERN.

### 5.2 Micro-mechanics

Smaller beam sizes in modern accelerators drive the technology for instrumentation. Wire grids (SEM grids are widely used to give direct measurements of beam profile as well as position and relative intensity (see Fig. 8). Instrument design has exploited technology for printed circuit boards (PCBs) and has seen significant improvements for recent machines such as LINAC4 at CERN. Ceramic supports are sputtered with a copper layer which is then locally removed by a lithographic and etching process to produce the connection structure for the wires. Wires are then typically attached either with a conductive glue or with 'bump-bonded' solder connections.

These instruments typically require a significant investment in tooling as well as advanced technical skills [8].

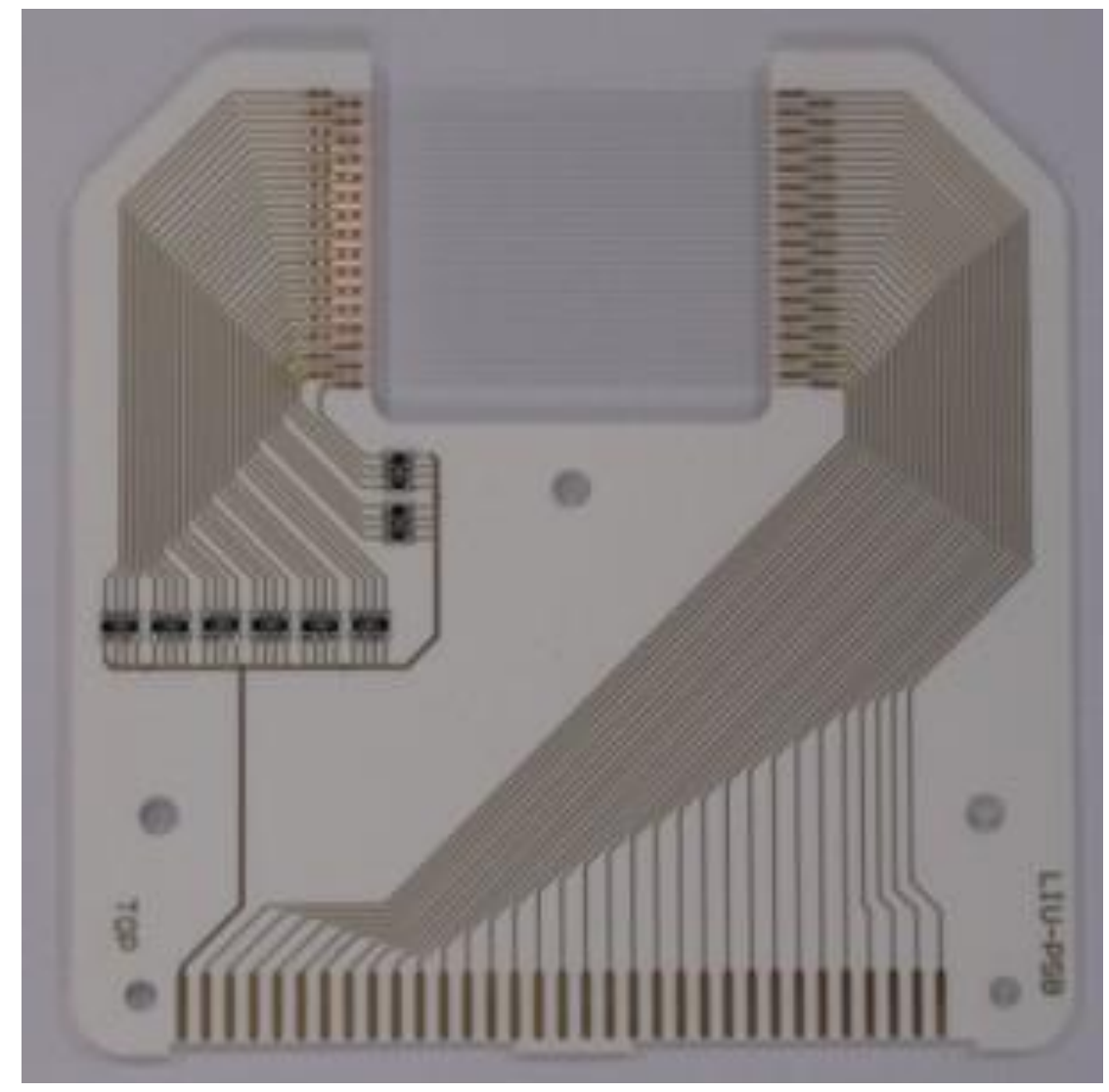

**Fig. 8:** SEM grid micro-mechanics.

### 5.3 Additive manufacturing (AM)

Additive manufacturing is a technology well-suited to beam instrumentation, allowing for rapid, small-series design and production. Development was made with titanium structures for the recent LIU fast wire scanner forks (see Figs. 9 and 10). A structure was topologically optimised for a specific stiffness and low inertial mass using a FE code. Parts were manufactured by Selective Laser Sintering (SLS) and then tested for outgassing in a UHV environment. Current limitations on production tolerances and feature size were overcome by post-machining.

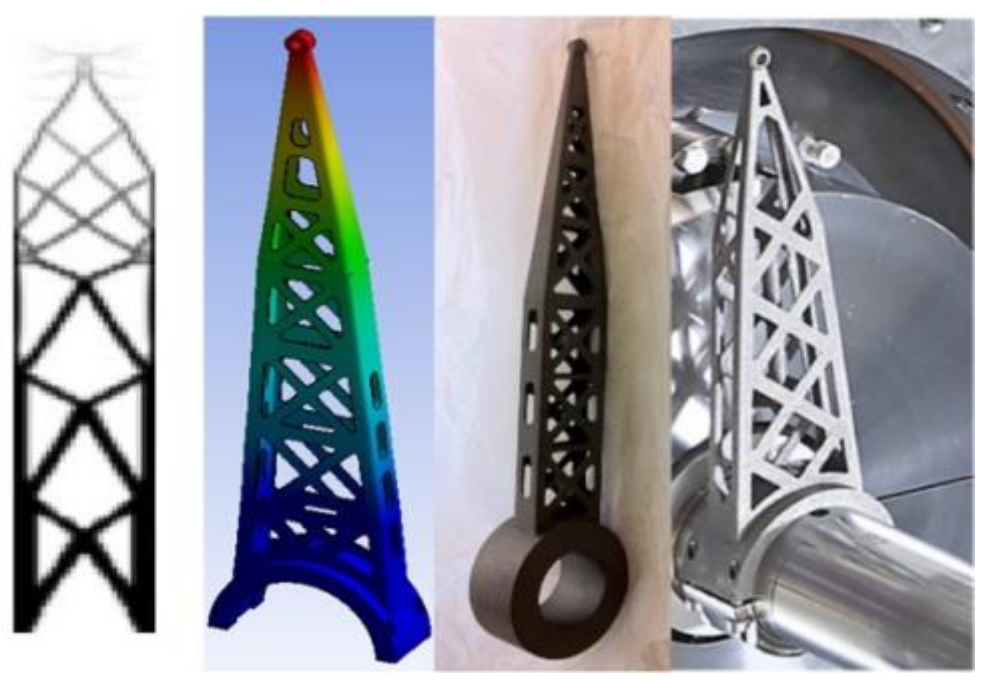

**Fig. 9:** Design of a wire scanner fork design – topological optimisation, finite element analysis, additive manufacturing.

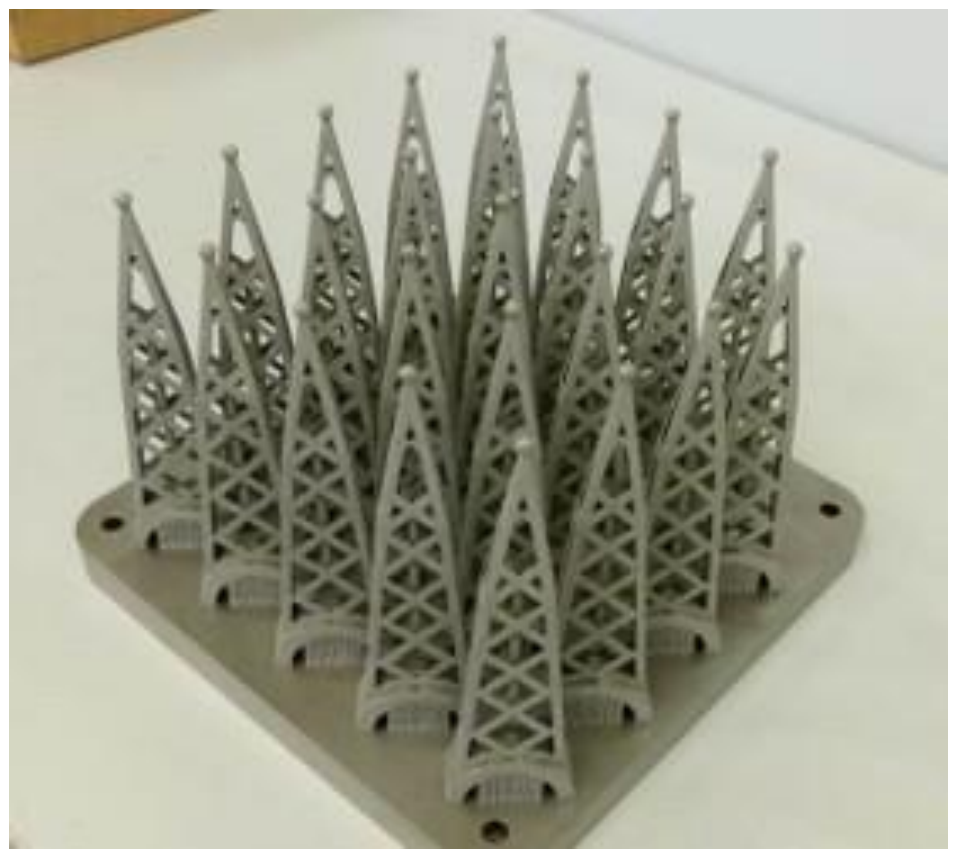

**Fig. 10:** Series production of wire scanner forks using additive manufacturing.

## 6 Putting it together: Examples of beam instruments

A number of examples or recent instruments are shown which combine recent technology in beam instrumentation as well as highlighting limitations.

### 6.1 Fast beam wire scanner

Figure 1 shows a fast wire scanner recently developed for the LHC Injector Upgrade programme. It is currently installed in the PSB, PS and SPS machines at CERN as well as in the ESS in Sweden.

A fine wire (currently 34 µm carbon fibre, but with upgrade plans for CNTs as per Section 5.1) is passed across the machine (see Section 4.1.2) at speeds upto 20 $ms^{-1}$. The wire is rigidly held in AM-produced titanium 'forks' as per Section 5.3. The forks are rotated on a shaft driven by a PMSM where rotor and stator are separated by a thin-walled vacuum chamber, magnetically coupled as per Section 3.1. Wire position is measured with an aluminium optical disc laser-etched with a micron-scale graticule. This is read by an ex-vacuum laser through an optical viewport as per Section 3.3.

This instrument has been very successful, replacing a number of obsolete instruments with significant increase in both performance and reliability. Issues of beam-coupled impedance heating (as per Section 4.2) were experienced in the SPS, which has the highest total intensity and energy of the chain. These resulted in systematic wire failures under certain conditions. A task force was established to understand and mitigate the issue with a number of studies over a few months [8] resulting in the retro-fitting of carefully designed ferrite blocks to detune a resonance and absorb energy.

### 6.2 SEM grid

SEM grids share the same material interaction challenges as scanners. In addition, specifications are often 'as many wires as possible, with as small a diameter as possible and with a separation 'as small as possible' to maximise the resolution. This pushes the boundaries in micro-mechanics as per Section 5.2. They are often used as an 'in-out' instrument, inserted for machine set-up and then removed with a transmission as per Section 3.1. This means that only the 'in' position needs to be precise and is normally defined with a mechanical end-stop with position surveyed before installation.

The many channels (upto 64) of independently conducting wires require multi-channel flexible cables and vacuum feedthroughs. Recent developments in vacuum-compatible polymers, insulators and connection technology [8] have allowed for applications in machines with more demanding vacuum requirements (see Fig. 11).

Despite significant improvements in automation of processes and new technology from PCB manufacture, these instruments remain a speciality, requiring expert design and construction with significant investment in tooling.

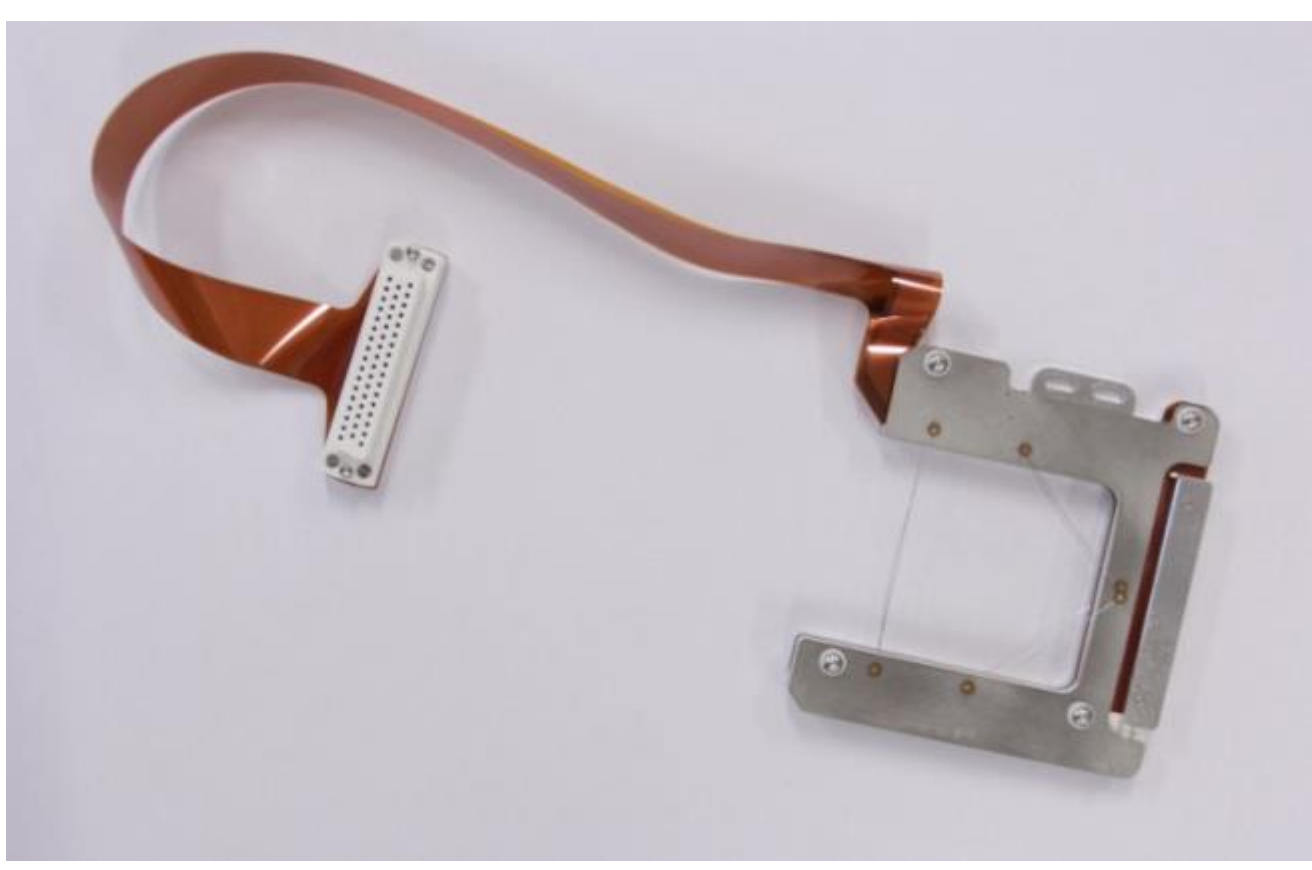

**Fig. 11:** SEM grid assembly for the nToF experiment at CERN showing grid, flex connector and multi-pin feedthrough.

## 7 Future perspectives

Although beam instrumentation is a core part of every accelerator, the state-of-the-art is driven largely by requirements from next-generation synchrotron light sources for very small, high brightness beams [10] and for energy frontier particle physics designs such as FCC, CECP, ILC and CLIC.

These are all lepton (electron) accelerators with common parameters of very small and high intensity beams, often with very short bunch lengths. Key challenges for instrumentation are therefore to improve resolution of position measurement and measure transverse and longitudinal profiles. The very high damage potential of these new machines drives a wider use of beam-loss monitoring as well as non-invasive measurement techniques using beam-gas interactions, synchrotron radiation and electro-magnetic effects.

Instrumentation mechanics responds to these challenges with research on new materials with low beam interactions as well as applying technology developed for micro- and nano- structures in electronics and telecommunications.

## 8 Concluding remarks

Beam instrumentation is a very diverse field with requirements driving a wide variety of measurables, technologies and materials with instrument sizes ranging from microns to 10's of meters. It is by definition a collaborative field as no one person can master all of the skills and competencies required. In addition. instruments typically require extensive research and development for one-off or small production series so tends to favour collaboration between universities and institutes rather than industry.

It is a field that is driven to innovate by requirements from future machines: "*An accelerator is only as good as what you can measure with it's instrumentation*", but also by the demands of operating and optimising existing machines: "*If you don't know what's wrong, you probably need a new instrument*".

## Acknowledgement

The author would like to acknowledge the many colleagues in beam instrumentation around the world and in particular in the Mechanics and Logistics Section of the CERN BI group.